\documentclass[%
 reprint,
superscriptaddress,
showpacs,
 amsmath,amssymb,
 aps,prl,
floatfix,
]{revtex4-1}

\usepackage{graphicx}
\usepackage{dcolumn}
\usepackage{bm}
\usepackage{braket}
\usepackage{color}
\usepackage{subcaption}
\usepackage[font=footnotesize,labelfont=bf]{caption}
\usepackage[font=footnotesize,labelfont=bf]{subcaption}
\usepackage{hyperref}

\begin{document}

\preprint{APS/123-QED}

\title{Rapid production of many-body entanglement in spin-1 atoms via cavity output photon counting}

\author{Stuart J. Masson}\email{smas176@aucklanduni.ac.nz} 
\affiliation{Dodd-Walls Centre for Photonic and Quantum Technologies, New Zealand}
\affiliation{Department of Physics, University of Auckland, Private Bag 92109, Auckland, New Zealand}

\author{Scott Parkins}\email{s.parkins@auckland.ac.nz}
\affiliation{Dodd-Walls Centre for Photonic and Quantum Technologies, New Zealand}
\affiliation{Department of Physics, University of Auckland, Private Bag 92109, Auckland, New Zealand}

\date{\today}

\begin{abstract}
We propose a simple and efficient method for generating metrologically useful quantum entanglement in an ensemble of spin-1 atoms that interacts with a high-finesse optical cavity mode. It requires straightforward preparation of $N$ atoms in the $m_F=0$ sublevel, tailoring of the atom-field interaction to give an effective Tavis-Cummings model for the collective spin-1 ensemble, and a photon counting measurement on the cavity output field. The photon number provides a projective measurement of the collective spin length $S$, which, for the chosen initial state, is heavily weighted around values $S\simeq\sqrt{N}$, for which the corresponding spin states are strongly entangled and exhibit Heisenberg scaling of the metrological sensitivity with $N$, as quantified by the quantum Fisher information.

\end{abstract}

\maketitle

Entanglement is a fundamental property of quantum mechanics. Two-body or two-mode entanglement is now readily producible and well studied, but study of many-body entangled systems and routine production of large many-body entangled ensembles are still open problems. The generation of such states is of interest not only to fundamental science, but for the use of such states as a resource for quantum information tasks and quantum metrology. In this latter context, there has been significant progress in the production of spin squeezing \cite{Ma11,Pezze18}. For $N$ particles in a non-correlated ensemble, the variance on measurements is limited by the standard quantum limit (SQL), which scales like $1/N$. In spin squeezing, entanglement is induced in an ensemble of atomic spins such that measurements of classical properties can be done more precisely. The fundamental limit on measurements with an ensemble  \emph{allowing for entanglement} is the Heisenberg limit with minimum variance scaling like $1/N^2$. 

A wide variety of spin squeezing techniques have been used to show sub-SQL variances. A common method involves the ``one-axis twisting'' mechanism 
\cite{Kitagawa93,Gross10,Leroux10PRL1,Riedel10,Ockeloen13,Muessel14}. Other procedures have produced up to a 100-fold reduction in the spin variance compared to classical states \cite{Hosten16}. These states have also been used for proof-of-principle, quantum enhanced implementations of atomic clocks \cite{Leroux10PRL2,Kruse16} and magnetometers \cite{Sewell12,Muessel14}, and to measure microwave fields \cite{Ockeloen13}. 
Other proposals, such as the two-axis counter-twisting scheme \cite{Kitagawa93}, offer a route to achieving Heisenberg limited metrological sensitivity, but these are yet to be implemented experimentally.

The present work concerns entanglement in an ensemble of spin-1 atoms, which, compared with spin-1/2 atoms, clearly require more degrees of freedom to describe, but concomitantly offer more degrees of freedom to entangle \cite{Sau10,Vitagliano11,Vitagliano14}. 
Indeed, proposals \cite{Duan02,Zhang13,HuangY15,Masson17,Sun17,Feldmann18,Kajtoch18} and experiments \cite{Bookjans11PRL1,Lucke11,Gross11,Hamley12,Hoang13,Peise15,Hoang16NatComm,Hoang16PNAS,Linnemann16,Kruse16,Luo17,Huang17} with spinor Bose-Einstein condensates (BECs) predict or have produced entanglement either on the Bloch sphere (e.g., squeezing in one of $\hat{S}_x$, $\hat{S}_y$, or $\hat{S}_z$, where $\hat{S}_i$ is the $i$-component of the collective atomic spin operator) or in the additional spinor degrees of freedom.

The metrological sensitivity of a quantum state can be captured by the quantum Fisher information (QFI). The variance of a measured phase $\theta$ imprinted by a classical parameter is bounded by $(\Delta \theta)^2 \geq \mathcal{F}^{-1}$, where $\mathcal{F}$ is the QFI. As such, the SQL states that for an optimal classical state the QFI scales as $N$ while the Heisenberg limit is signified by a QFI that scales like $N^2$. For pure states, the QFI over some generator $\hat{G}$ is $\mathcal{F} = 4 (\Delta \hat{G})^2$. More generally, for a density matrix $\rho$, decomposed into eigenstates as $\rho = \sum_i \xi_i\ket{e_i}\bra{e_i}$, the QFI is given by
\begin{equation}
\mathcal{F} = 2 \sum\limits_{i,j} \frac{(\xi_i - \xi_j)^2}{\xi_i+\xi_j} |\bra{e_i} \hat{G} \ket{e_j}|^2.
\end{equation}

Typically these quantities would be maximised over a set of generators to find the best possible QFI. In this work, given that the state generation protocol we propose produces varying, heralded states, we choose a single generator to consider: $\hat{Q}_{xx} - \hat{Q}_{yy}$. Here, $\mathbf{\hat{Q}}$ is the nematic tensor operator and $\hat{Q}_{ij} = \sum_{n=1}^N \hat{S}_i^{(n)} \hat{S}_j^{(n)} + \hat{S}_j^{(n)} \hat{S}_i^{(n)} - (4/3) \delta_{ij}$ \cite{Hamley12,HuangY15,Masson17}, where $i,j \in \set{x,y,z}$, $\hat{S}_i^{(n)}$ are spin-1 angular momentum operators for a single atom, and $\delta_{ij}$ is the Kronecker delta function. This generator, in a bosonic mode operator picture where $\hat{b}_i (\hat{b}^\dagger_i)$ is the annihilation (creation) operator for a particle in state $\ket{m_F=i}$, is given by $2(\hat{b}^\dagger_{+1}\hat{b}_{-1} + \hat{b}^\dagger_{-1}\hat{b}_{+1})$, and so involves a transfer of atoms between the $\ket{m_F=\pm1}$ states. If we were to consider only these two states, reducing the atoms to effective two-level systems, then the algebra would give this as $2\hat{S}_x$.

In this Letter, we propose a new method to produce entanglement in an ensemble of spin-1 atoms. We use interactions mediated by cavity-assisted Raman transitions, building on previous work for generating such interactions with two-level (spin-1/2) atoms \cite{Dimer07,Morrison08PRL,Morrison08PRA,Zhiqiang18}. This approach has previously been followed to produce an effective Dicke model \cite{Zhiqiang17,Masson17} and spin-exchange interactions \cite{Davis18,Marino18,Masson18Singlet} for spinor (spin$\,\geq 1$) atoms. Here, we engineer instead an effective Tavis-Cummings (TC) model for an ensemble of spin-1 atoms, which, as we show, can be used to herald, via a photon counting measurement on the cavity output field, the production of one of a family of highly entangled, many-body quantum states. We show further that the average result of this procedure, for ideal photon detection, in fact gives Heisenberg scaling of the QFI, while for non-ideal photon detection, the method still retains metrological sensitivity beyond the SQL and with scaling significantly better than linear. We also show that by alternating between TC and anti-Tavis-Cummings (anti-TC) interactions, so as to produce a sequence of cavity output pulses and corresponding photon counting measurements, it is in principle possible to regain Heisenberg scaling even with finite detection efficiency.

\begin{figure}[b]
\includegraphics[width=0.35\textwidth]{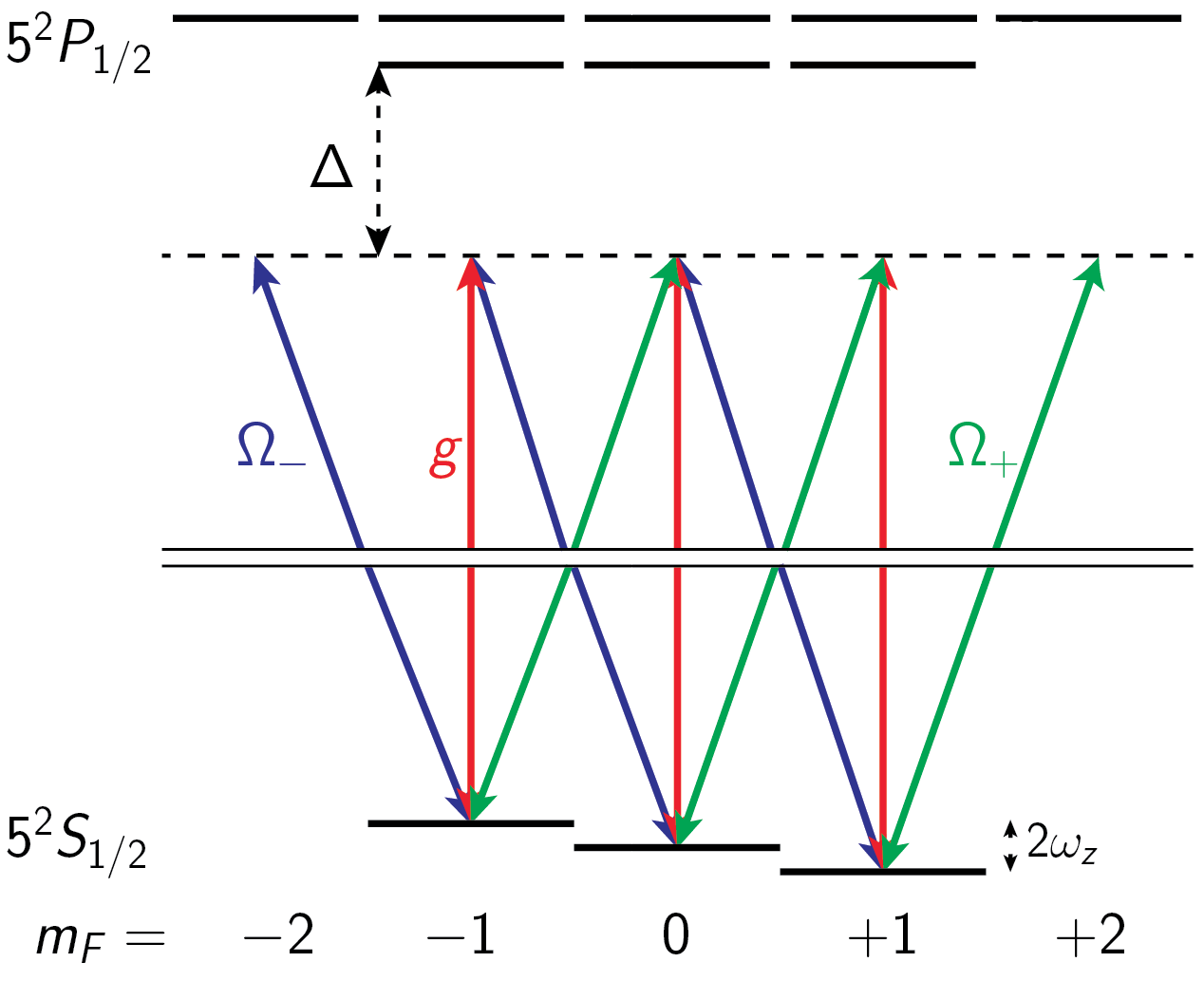}
\caption{Implementation of an effective Dicke model using the $F=1$ ground state of $^{87}$Rb. Interactions are mediated by detuned Raman transitions on the $D_1$ line mediated by a cavity mode (red) and $\sigma_-$ (blue) and $\sigma_+$ (green) polarised lasers.\label{leveldiagram}}
\end{figure}

For a specific system, we consider $N$ $^{87}$Rb atoms confined tightly within an optical cavity and pumped into the $F=1$ ground hyperfine level \footnote{Note that the scheme in this letter works equally well with pumping into the $F=2$ level, in which case we realize an ensemble of spin-2 atoms. With, e.g., cesium, one could similarly realize the scheme with spin-3 or spin-4 atoms \cite{Masson17}.}.
As shown in Fig.~\ref{leveldiagram}, we use a scheme of cavity-assisted Raman transitions on the $D_1$ line to introduce effective interactions between the atoms and the cavity mode. With both lasers on and detuning $\Delta$ much larger than the width of the excited state hyperfine structure, the model of the system reduces to an effective, dissipative Dicke model for the cavity mode and an atomic spin-1 ensemble ($\hbar =1$),
\begin{equation}\label{eq:ME}
\dot{\rho} = -i[\hat{H},\rho] + \kappa (2\hat{a}\rho \hat{a}^\dagger - \hat{a}^\dagger \hat{a} \rho - \rho \hat{a}^\dagger \hat{a})
\end{equation}
where $\rho$ is the density operator for the composite atom-cavity system, $\kappa$ is the cavity field decay rate, $\hat{a}$ is the cavity mode annihilation operator, and
\begin{align}\label{dickemodel}
\notag\hat{H} = &\omega \hat{a}^\dagger \hat{a} + \omega_0 \hat{S}_z \\&+ \lambda_- (\hat{a} \hat{S}_+ + \hat{a}^\dagger \hat{S}_-) + \lambda_+ (\hat{a} \hat{S}_- + \hat{a}^\dagger \hat{S}_+), 
\end{align}
where we have introduced collective spin operators $\hat{S}_{z,\pm}$, which are sums of $N$ spin-1 operators. The coefficients in (\ref{dickemodel}) are given by
\begin{align}
\omega &= \omega_c - \frac{\omega_- + \omega_+}{2} + \frac{Ng^2}{3\Delta} , \\
\omega_0 &= \omega_z - \frac{\omega_- - \omega_+}{2} + \frac{\Omega_-^2 - \Omega_+^2}{24\Delta}, \\
\lambda_\pm &= \frac{g\Omega_\pm}{12\sqrt{2}\Delta}.
\end{align}
Here $\omega_c$ is the frequency of the cavity mode, $\omega_\pm$ and $\Omega_\pm$ are the bare and Rabi frequencies, respectively, of the $\sigma_\pm$ polarised laser fields, $\omega_z$ is the Zeeman splitting of the $F=1$ levels, $g$ is the single-atom-cavity coupling strength (for the $^{87}$Rb $D_2$ line cycling transition), and $\Delta$ is the detuning of the fields from the atomic resonance.

\begin{figure}[t]
\includegraphics[width=0.5\textwidth]{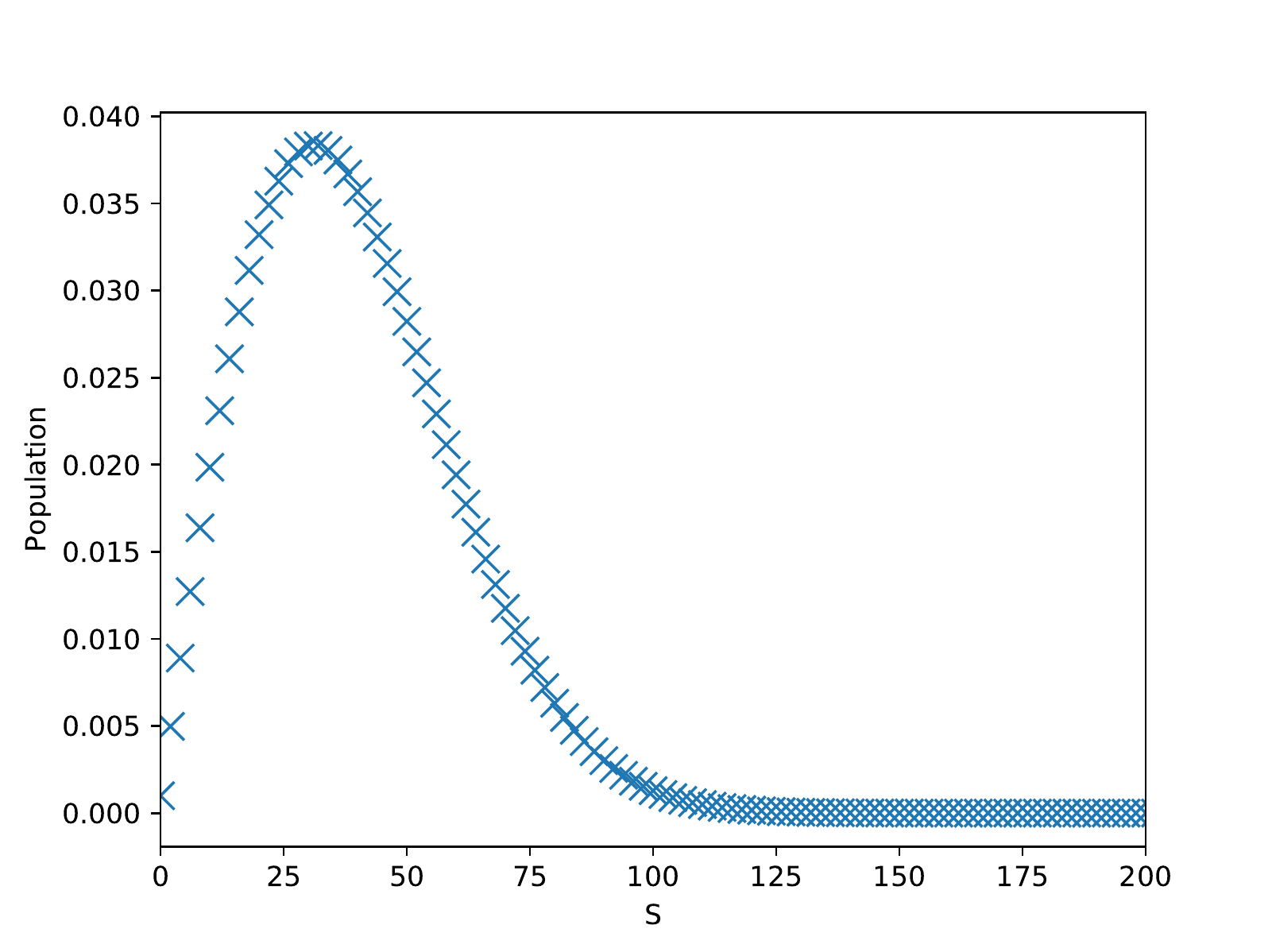}
\caption{Populations $|c_S|^2$ for even $S$ in the collective Dicke states $\ket{S,0}$ for the product state $\ket{m_F=0}^{\otimes N}$ of $N=1000$ spin-1 atoms. Populations for states beyond $S=200$ are not shown as the total sum of these populations is $\sum_{S>200}|c_S|^2=1.34\times10^{-9}$.\label{superpositioncoefficients}}
\end{figure}

We consider this system with an initial atomic state $\ket{m_F=0}^{\otimes N}$. This state does not have a certain spin length. Rather, it is given by a superposition of states of different spin lengths, which, in a representation of Dicke states $\ket{S,0}$, can be written
\begin{equation}\label{eq:psi0}
\ket{m_F=0}^{\otimes N} = \sum\limits_{S= 0}^N c_S \ket{S,0}.
\end{equation} 
For even numbers of atoms, $c_S = 0$ for all odd $S$. Odd numbers of atoms instead have $c_S=0$ for even $S$.

We build this superposition by using the Racah formula, which for $\ket{S,0}\otimes\ket{1,0}$ reduces to
\begin{equation}
\ket{S,0} \otimes\ket{1,0} = \sqrt{\frac{S+1}{2S+1}} \ket{S+1,0} - \sqrt{\frac{S}{2S+1}} \ket{S-1,0}.
\end{equation}
We calculate the coefficients $\{ c_S\}$ of the superposition in (\ref{eq:psi0}) by iterating this formula $N-1$ times. An example of the resulting distribution of $|c_S|^2$ values is shown in Fig.~\ref{superpositioncoefficients} for $N=1000$. One sees that the dominant constituents of the state actually have much shorter spin length than the maximum possible value of $S=N$, with the peak of the distribution centered at $S\simeq\sqrt{N}$.

It can be shown that all states with a definite spin length $S<N$ are entangled \cite{Vitagliano11,Vitagliano14}. This means that the individual elements of the initial superposition (\ref{eq:psi0}) are \textit{on their own} entangled, though the superposition of them is not. Our proposal is thus to project out one element of the superposition and so generate entanglement in the ensemble.

\begin{figure}[t]
\includegraphics[width=0.5\textwidth]{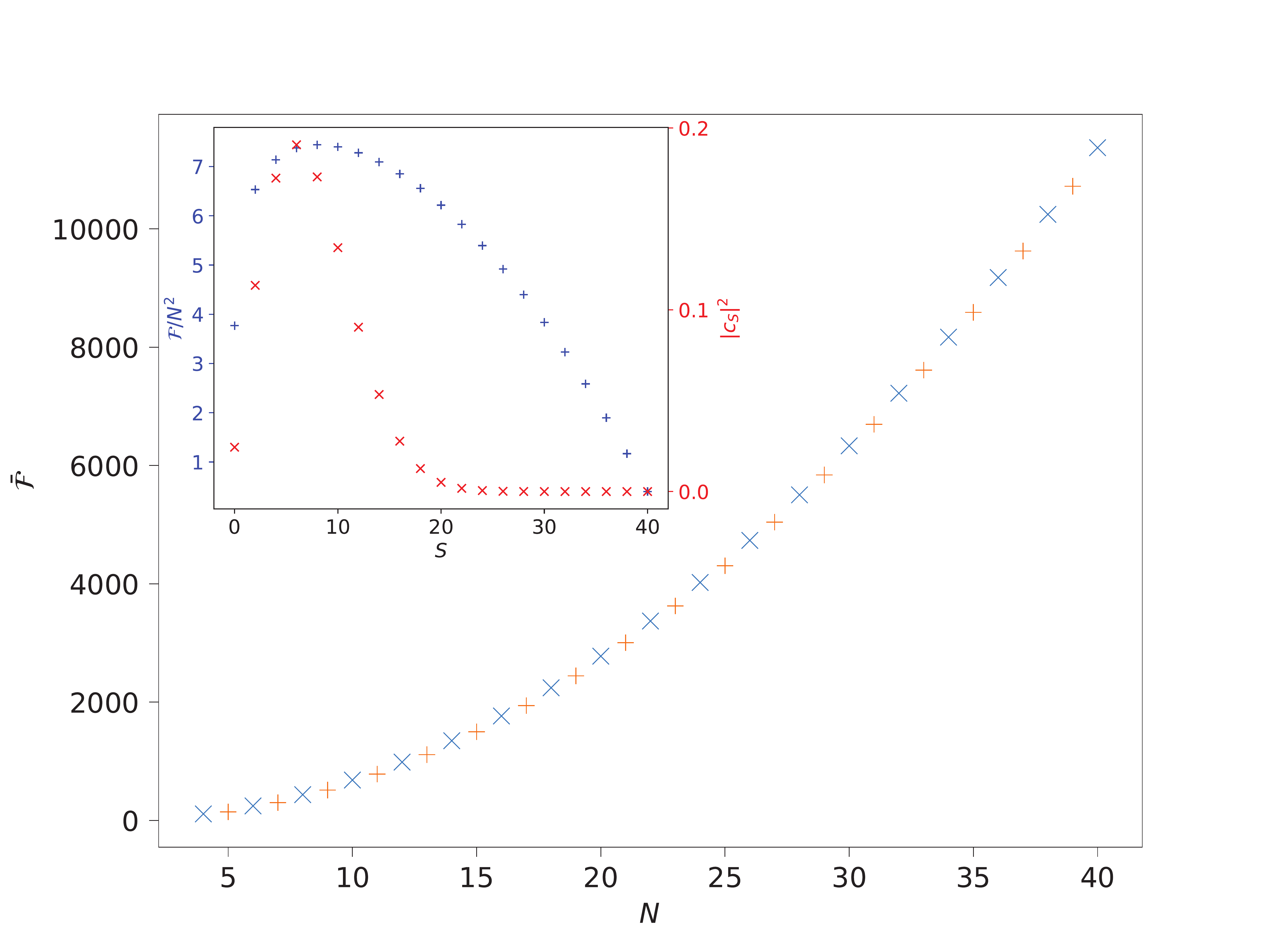}
\caption{Average QFI on the generator $\hat{Q}_{xx}-\hat{Q}_{yy}$ of states $\ket{S,-S}$ weighted by initial populations $|c_S|^2$ of the states $\ket{S,0}$ in (\ref{eq:psi0}). Even and odd numbers are represented differently due to their slightly different dependence on $N$. The inset shows the populations and QFI of the individual states $\ket{S,-S}$ for $N=40$.\label{perfectdetectorqfi}}
\end{figure}

To do this, we consider (\ref{dickemodel}) with $\lambda_+ = 0$, which reduces the Hamiltonian to a TC model for the collective spin-1 ensemble,
$\hat{H} =  \lambda_- (\hat{a} \hat{S}_+ + \hat{a}^\dagger \hat{S}_-)$, where we further assume that $\omega$ and $\omega_0$ can be set to $\omega \simeq \omega_0 \simeq 0$. Since the system is open, due to cavity loss at rate $\kappa$, any initial state $\ket{S,0}\otimes\ket{0}_{\rm cav}$ (where $\ket{0}_{\rm cav}$ denotes the vacuum state of the cavity mode) will evolve, subject to (\ref{eq:ME}) with the TC Hamiltonian, to the unique steady state $\ket{S,-S}\otimes\ket{0}_{\rm cav}$, with emission from the cavity of a pulse containing precisely $S$ photons. For $\sqrt{S}\lambda_-\lesssim\kappa/2$, the duration of this pulse is $t_{\rm pulse}\simeq (S\lambda_-^2/\kappa)^{-1}$, while for $\sqrt{S}\lambda_->\kappa/2$ the timescale is set simply by $\kappa$ (i.e., a few times $1/\kappa$).

It follows that, if a photon counting measurement is performed on the cavity output field with an ideal photodetector, then the system will be projected into a particular state $\ket{S,-S}$ from the initial superposition (\ref{eq:psi0}) with probability $|c_S|^2$. Given the strong weighting of the distribution $|c_S|^2$ towards values $S\sim\sqrt{N}$ (Fig.~\ref{superpositioncoefficients}), the efficiency (and simplicity) of this procedure for generating entangled spin states is clear.

To calculate the average entanglement this process introduces in the ensemble, we consider the average QFI of a single run. For a perfect photodetector this is simply
\begin{equation}
\bar{\mathcal{F}} = \sum\limits_{S=0}^N |c_S|^2 \mathcal{F}(\ket{S,-S}) ,
\end{equation}
where $\mathcal{F}(\ket{S,-S})$ is the QFI of state $\ket{S,-S}$ with respect to the generator $\hat{Q}_{xx} - \hat{Q}_{yy}$.
Fig.~\ref{perfectdetectorqfi} shows that this quantity increases with $N$ in a quadratic fashion. For even $N$, a fit of the data gives the average QFI as $\bar{\mathcal{F}} = 6.52N^{2.02}$. For odd $N$ we find $\bar{\mathcal{F}} = 5.17N^{2.09}$. 
The fits imply a slightly better than quadratic scaling, but we believe that this is due to the contributions of lower order terms; for sufficiently large $N$ these should be negligible and the scaling should return to being purely quadratic. So, we find that our heralded state has optimal scaling for quantum metrology on the generator $\hat{Q}_{xx} - \hat{Q}_{yy}$. In fact, there are a range of generators for which we can show quadratic scaling. These generators are all higher order operators than the angular momentum operators, showing that this entanglement is a distinctly spinor phenomenon.

\begin{figure}[b]
\includegraphics[width=0.5\textwidth]{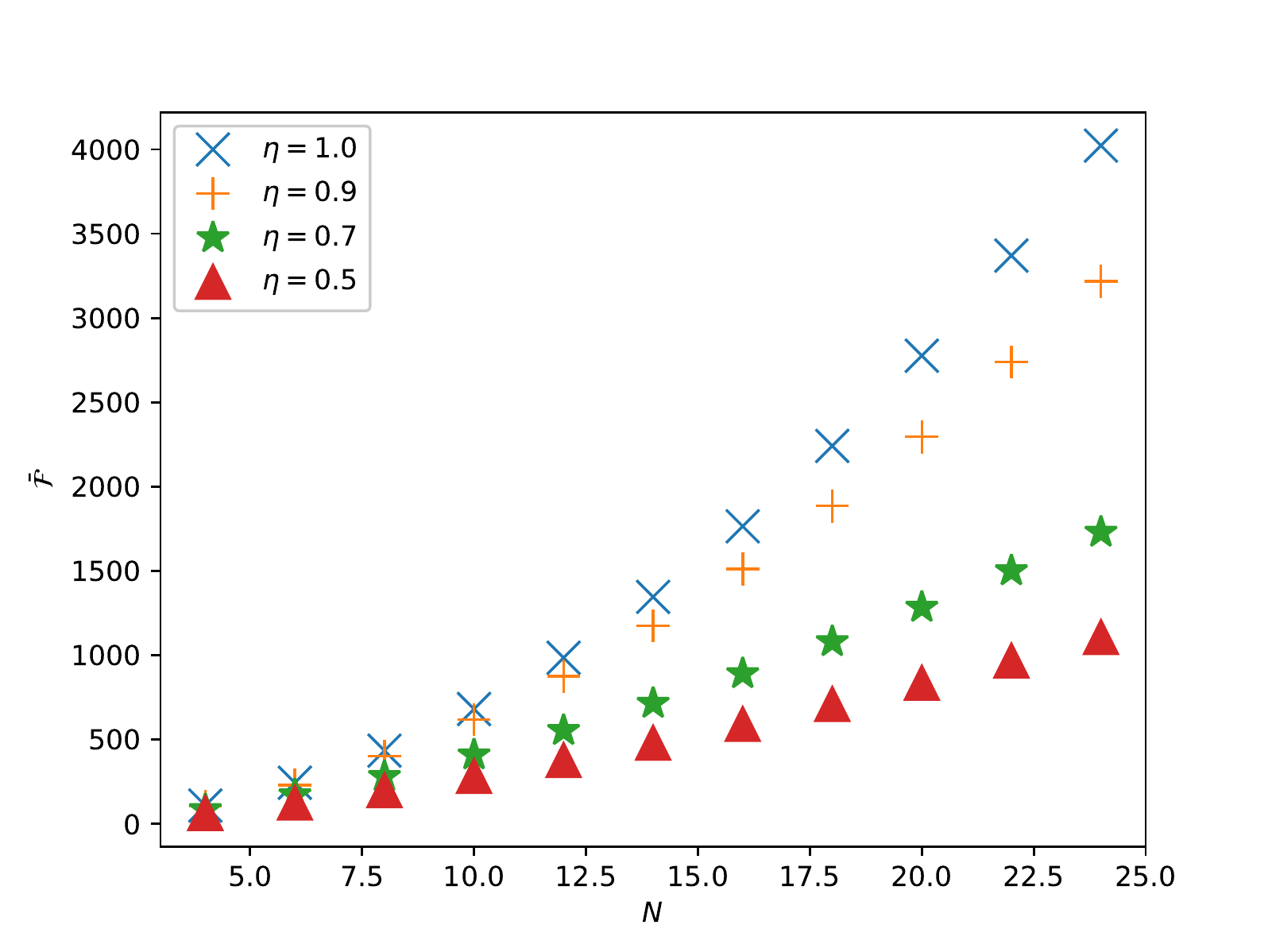}
\caption{Average QFI on the generator $\hat{Q}_{xx}-\hat{Q}_{yy}$ of states $\rho_n$ weighted by the probability $p(n)$ of measuring $n$ photons, as given by (\ref{eq:Fbar}) and (\ref{eq:pn}), for various photodetector efficiencies $\eta$. \label{imperfectscaling}}
\end{figure}

Now consider the more realistic case in which we have a detector of finite photon detection efficiency $\eta$. The state resulting from our measurement scheme can then be modeled as a mixed state, i.e., given the photodetector records $n$ photons in a single run, the resultant state can be written as
\begin{align} \label{onepulseequation}
& \rho_n = \sum\limits_{S\geq n}^{N} p(S|n) \ket{S,-S}\bra{S,-S} ,
\\
\mathrm{with} \;\;\;\; & p(S|n) = \frac{|c_S|^2 \begin{pmatrix} S \\ n \end{pmatrix} \eta^n (1-\eta)^{S-n}}{\sum\limits_{k\geq n}^{N} |c_k|^2 \begin{pmatrix} k \\ n \end{pmatrix} \eta^n (1-\eta)^{k-n}} ,
\end{align}
where the sum is only over states that can produce $n$ or more photons.

Actually, this is somewhat of a simplification, as the times at which the photons are detected could in principle provide extra information related to the likelihood of each state. We choose to ignore this aspect of the detection process, but note that, since this information would improve knowledge of the state, using it would only enhance our scheme.
We also ignore a possible dark count rate for the photodetector. However, this could be included by assigning a finite probability to the possibility of  detection events being the result of dark counts. 

The states (\ref{onepulseequation}) are not perfect projections, but they do have a reduced width in $S$ and, for $N\gg 1$, are entangled with virtual certainty, as only the state $\ket{N,-N}$ does not feature entanglement and $|c_N|^2\sim 2^{-N}\simeq 0$.
In other words, even with finite photodetector efficiency, entanglement is still produced with essentially unit efficiency.

As with the perfect detector, we can consider an average QFI where now
\begin{align}\label{eq:Fbar}
& \bar{\mathcal{F}} = \sum\limits_{n=0}^N p(n) \mathcal{F}(\rho_n) ,
\\
\mathrm{with} \;\;\;\; p(n) & = \sum\limits_{S\geq n}^{N} |c_S|^2 \begin{pmatrix} S \\ n \end{pmatrix} \eta^n (1-\eta)^{S-n}.  \label{eq:pn}
\end{align}
This average is shown in Fig.~\ref{imperfectscaling}. The scaling of the QFI is still better than linear, but it is no longer quadratic. Nevertheless, for $\eta = 0.9$ the data is fitted by $\bar{\mathcal{F}} = 7.56N^{1.91}$, while for $\eta = 0.5$ the scaling is still $\sim N^{1.58}$.

\begin{figure}[b]
\includegraphics[width=0.5\textwidth]{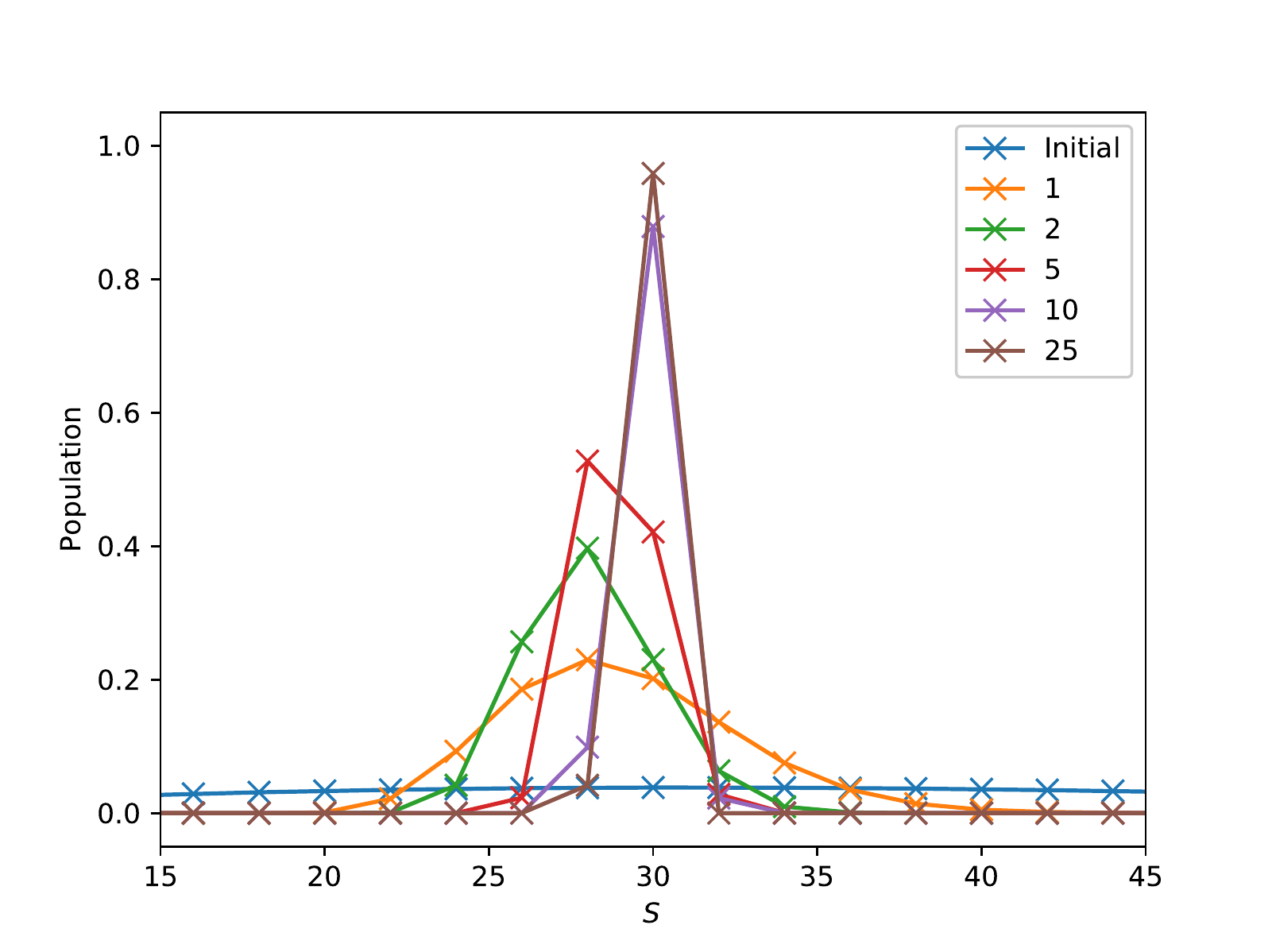}
\caption{Population as a function of spin length $S$ for a varying total number of output (multiphoton) pulses generated by a sequence of TC and anti-TC interactions. 
Each pulse is simulated with a binomially-distributed registered photon number using a detection efficiency of $\eta=0.7$ and an assumed actual photon number based on a spin length $S=30$.\label{imperfectmultiple}}
\end{figure}

Significantly, using an imperfect detector does not in fact rule out the possibility of Heisenberg scaling. The flexibility of our engineered atom-cavity interaction offers a straightforward means of improving our knowledge of the spin length. Following the first output pulse of photons resulting from the effective TC interaction, one can switch the polarization of the laser field such that, in model (\ref{dickemodel}), one now has $\lambda_- = 0$ and $\lambda_+ \neq 0$, corresponding to an anti-TC model. The steady states are now $\ket{S,+S}\otimes\ket{0}_{\rm cav}$, and the resulting transfer $\ket{S,-S} \rightarrow \ket{S,+S}$ will produce an output pulse of $2S$ photons. Detection of this pulse provides a second measurement and subsequent, further narrowing of the distribution in $S$.

\begin{figure}[t]
\includegraphics[width=0.5\textwidth]{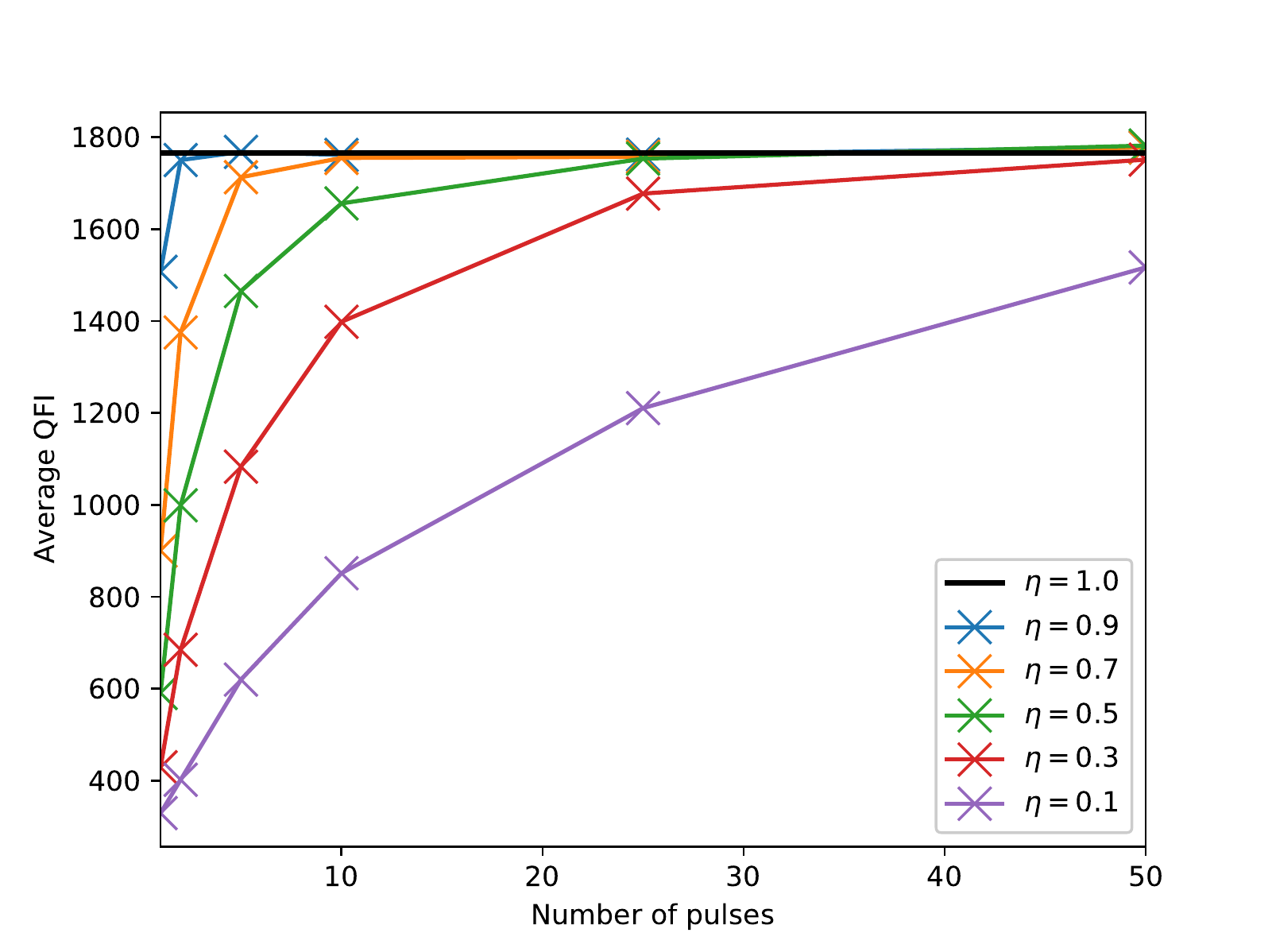}
\caption{Sample mean QFI on generator $\hat{Q}_{xx} - \hat{Q}_{yy}$, for $N=16$, of 1000 states produced by randomly selecting a spin length $S$ and a sequence of binomially-distributed measured photons given the initial state and a photon detection efficiency $\eta$ respectively. \label{imperfectmultiplescaling}}
\end{figure}

In fact, we can consider a sequence of alternating TC and anti-TC interactions, producing a corresponding sequence of pulses and measurements $\{ n\} =\{ n,n_1,n_2,...,n_{i-1}\}$. The density matrix conditioned upon a further measurement of $n_i$ photons can be written
\begin{equation}
\rho_{\set{n},n_i} = \sum\limits_{S\geq S_{\mathrm{min}}}^N p(S|\set{n},n_i) \ket{S,\pm S}\bra{S,\pm S} ,
\end{equation}
with probabilities
\begin{equation}
p(S|\set{n},n_i) = \frac{p(S|\set{n}) \begin{pmatrix} 2S \\ n_i \end{pmatrix} \eta^{n_i} (1-\eta)^{S-n_i}}{\sum\limits_{k\geq S_{\mathrm{min}}}^{N} p(S|\set{n}) \begin{pmatrix} 2k \\ n_i \end{pmatrix} \eta^{n_i} (1-\eta)^{k-n_i}} ,
\end{equation}
where $S_{\mathrm{min}}$ is the largest value in the set $\set{n,n_1/2,\dots,n_i/2}$. That is, we iteratively produce a state conditioned on a sequence of binomially-distributed photon numbers.

A numerical example of such a sequence is shown in Fig.~\ref{imperfectmultiple} and it clearly illustrates that with each measurement we gain more knowledge about the state, narrowing the distribution in $S$. After enough polarization switches and output pulses we have, with almost certainty, projected out a state of definite spin length.

Taking a sampling approach, Fig.~\ref{imperfectmultiplescaling} shows that for lower efficiency more switches and their associated output pulses are necessary. However, eventually, a state of definite spin length is always generated, and so the average QFI simply reduces to the result for an ideal detector. This means that, in principle, we can achieve Heisenberg level scaling for the metrological sensitivity in spite of finite photodetector efficiency.

For our scheme, we note also that if both lasers are on ($\lambda_+=\lambda_->0$), then we realize an effective Dicke model, in which case the different $S$ states are heralded by the output photon {\it flux}. This flux could be sensitively measured through heterodyne detection, with longer averaging times providing the mechanism for narrowing the distribution in $S$.

Finally, we consider briefly some potential experimental parameters and timescales. Given, e.g., cavity QED parameters $\{ g,\kappa ,\gamma\}/(2\pi )=\{ 10,0.2,6\}\,$MHz and $N=10^4$ atoms, one finds $t_{\rm pulse}\simeq (S\lambda_-^2/\kappa)^{-1}\simeq 10\,\mu$s with the choice $\Omega_-/\Delta =0.01$, and setting $S=\sqrt{N}$ (which corresponds to the most probable spin length in the initial atomic state). Hence, the timescale for preparation of the entangled state is potentially very fast, and, indeed, orders of magnitude shorter than the characteristic timescales associated with the generation of entangled spin states via collisional dynamics or adiabatic ground state transformations in spin-1 BECs.


The authors acknowledge the contribution of NeSI high-performance computing facilities to the results of this research. New Zealand's national facilities are provided by the New Zealand eScience Infrastructure and funded jointly by NeSI's collaborator institutions and through the Ministry of Business, Innovation and Employment's Research Infrastructure program.


\bibliographystyle{apsrev4-1}

\end{document}